\documentclass[a4paper,11pt]{article}
\usepackage{pos}

\title{Towards glueball masses of large-$N$~$\mathrm{SU}(N)$ Yang--Mills theories without topological freezing via parallel tempering on boundary conditions}
\ShortTitle{Glueball masses at large-$N$ via parallel tempering on boundary conditions}

\author*[a]{Claudio Bonanno}
\author[b]{Massimo D'Elia}
\author[c]{Biagio Lucini}
\author[d,e]{Davide Vadacchino}

\affiliation[a]{INFN Sezione di Firenze, Via G.~Sansone 1, Sesto Fiorentino, I-50019 Firenze, Italy}
\affiliation[b]{Università di Pisa and INFN Sezione di Pisa, Largo B.~Pontecorvo 3, I-56127 Pisa, Italy}
\affiliation[c]{Department of Mathematics, Faculty of Science and Engineering, Swansea University, Fabian Way, Swansea, SA1 8EN, Wales, UK}
\affiliation[d]{School of Mathematics and Hamilton Mathematics Institute, Trinity College, Dublin 2, Ireland}
\affiliation[e]{Centre for Mathematical Sciences, University of Plymouth, Plymouth, PL4 8AA, UK}
\affiliation[]{}

\emailAdd{claudio.bonanno@fi.infn.it}
\emailAdd{massimo.delia@unipi.it}
\emailAdd{b.lucini@swansea.ac.uk}
\emailAdd{davide.vadacchino@plymouth.ac.uk}

\abstract{Standard local updating algorithms experience a critical slowing down close to the continuum limit, which is particularly severe for topological observables. In practice, the Markov chain tends to remain trapped in a fixed topological sector. This problem further worsens at large $N$, and is known as \emph{topological freezing}. To mitigate it, we adopt the parallel tempering on boundary conditions proposed by M.~Hasenbusch. This algorithm allows to obtain a reduction of the auto-correlation time of the topological charge up to several orders of magnitude. With this strategy we are able to provide the first computation of low-lying glueball masses at large $N$ free of any systematics related to topological freezing.}

\FullConference{The 39$^{\text{th}}$ International Symposium on Lattice Field Theory, LATTICE2022\\8$^{\text{th}}$-13$^{\text{th}}$ August, 2022\\Rheinische Friedrich-Wilhelms-Universität Bonn, Bonn, Germany}

\usepackage{braket}
\usepackage[shortlabels]{enumitem}
\usepackage{placeins}
\usepackage{makecell}

\newcommand{\beq}{\begin{eqnarray}}
\newcommand{\eeq}{\end{eqnarray}}
\newcommand{\beqnn}{\begin{eqnarray*}}
\newcommand{\eeqnn}{\end{eqnarray*}}

\newcommand{\Tr}{\mathrm{Tr}}
\newcommand{\SO}{\mathrm{SO}}
\newcommand{\SU}{\mathrm{SU}}
\newcommand{\CP}{\mathrm{CP}}
\newcommand{\PC}{\mathrm{PC}}
\newcommand{\R}{\mathrm{R}}

\begin{document}
\maketitle

\section{Introduction}

Glueballs, bound states of gluons only, are one of the few Standard Model predictions which still lack a satisfying experimental confirmation, since, despite several attempts, to date only one recent indirect evidence of their existence has been found at colliders experiments, see for example~\cite{TOTEM:2020zzr}. From the theoretical side, there has been a tremendous effort of the theoretical community to provide more and more precise estimations of glueballs masses, especially by means of lattice simulations~\cite{Lucini:2001ej,Lucini:2004my,Lucini:2010nv,Hong:2017suj,Bennett:2020hqd,Bennett:2020qtj,Athenodorou:2020ani,Athenodorou:2021qvs}, which are a natural tool to explore this topic, being the existence of glueball states a purely non-perturbative prediction stemming from the QCD confining properties.

Most lattice predictions have been obtained in pure-gauge Yang--Mills theories. The possibility of exploring the large-$N$ limit ($N=$~number of colors) is particularly attractive, as it provides a reasonable a reasonable approximation of QCD (being correction to $N=\infty$ suppressed as powers of $1/N^2$) and, at the same time, allows to simplify computations (the absence of quarks and the large-$N$ limit make all glueballs non-decaying and non-interacting).

Numerical methods to extract glueball masses from lattice gauge configurations have been refined tremendosuly in the last two decades~\cite{BERG1983109, APE:1987ehd, TEPER1987345, Morningstar:1997ff, Teper:1998te,Morningstar:1999rf, Lucini:2001ej, Lucini:2004my, Blossier:2009kd,Lucini:2010nv}, in particular concerning the control over systematic errors, which are by far the dominant source of uncertainties in these kind of calculations. However, a possible systematic which has not been satisfyingly checked in the literature is the impact of \emph{topological freezing} on glueball mass results.

When approaching the continuum limit, Monte Carlo Markov chains tend to remain trapped in a fixed topological sector if local updating algorithms are employed to explore the space of configurations~\cite{Alles:1996vn,DelDebbio:2004xh,Schaefer:2010hu}. This is due to the loss of effectiveness of local updating steps in decorrelating the configurations when the lattice spacing is small, thus prohibitively large statistics are required to generate a representative sample close to the continuum. The severity of such Critical Slowing Down increases exponentially with $N$~\cite{Lucini:2004yh,DelDebbio:2006yuf, Bonati:2016tvi,Bonati:2017woi,Bonanno:2018xtd,Berni:2019bch,Berni:2020ebn,Bennett:2022ftz,Bennett:2022gdz}, so that in practice the Monte Carlo evolution of the topological charge $Q$ already freezes on coarse lattices when $N$ is large. For this reason, most of the results for glueball masses at large $N$ have been obtained for frozen or nearly-frozen topology. Since on theoretical grounds we expect that computing a glueball mass in a fixed topological sector introduces non-trivial corrections~\cite{Brower:2003yx} with respect to the correct result (obtained averaging over all topological sectors), it is of the utmost importance to check that topological freezing does not result in an unwanted bias on the computed glueball masses, at least within the typical level of accuracy that can be reached with current state-of-the-art techniques.

In this paper, we report on the main results achieved in Ref.~\cite{Bonanno:2022yjr}, where the first computation of glueball masses at large $N$ free of the systematics effects of topological freezing has been obtained by means of the \emph{parallel tempering on boundary conditions} algorithm proposed by M.~Hasenbusch for $2d$ $\CP^{N-1}$ models~\cite{Hasenbusch:2017unr}. This algorithm was recently implemented for $4d$ $\SU(N)$ pure-gauge theories too~\cite{Bonanno:2020hht} and it dramatically mitigated the effects of topological freezing. As a matter of fact, such algorithm allows to reduce the auto-correlation time of $Q$ (the number of updating steps necessary to generate two decorrelated configurations with different topological charge) by up to more than two orders of magnitude at large $N$.

In Sec.~\ref{sec:setup} we outline our numerical setup; in Sec.~\ref{sec:results} we summarize results of~\cite{Bonanno:2022yjr} for low-lying glueball masses obtained for $\SU(6)$ for a lattice spacing $a\sim0.09$~fm with parallel tempering and we compare them with those obtained by standard algorithms in~\cite{Athenodorou:2021qvs}; in Sec.~\ref{sec:conclusions} we draw our conclusions.

\section{Lattice setup}\label{sec:setup}

\begin{itemize}
\item \emph{Parallel tempering on boundary conditions}

We simulate $N_r$ replicas of a $L^4$ lattice imposing periodic boundary conditions everywhere but on a small region, \emph{the defect}, where different boundary conditions are considered as a function of the replica $r$. Such different boundary conditions are imposed multiplying the gauge coupling of the links crossing the defect by a constant $0\le c(r) \le 1$, where the extremes are periodic ($c(0)=1$) and open boundaries ($c(N_r-1)=0$). The lattice action of a replica thus reads:
\beq
S_L^{(r)} = -\frac{\beta}{N} \sum_{x, \mu>\nu} K^{(r)}_{x,\mu} K^{(r)}_{x+\hat{\mu},\nu} K^{(r)}_{x+\hat{\nu},\mu} K^{(r)}_{x,\nu} \,\, \Re \Tr \Pi^{(r)}_{x,\mu\nu},
\eeq
with $\Pi_{\mu\nu}(x)$ the plaquette and $K^{(r)}_{x,\mu}=c(r)$ if link $(x,\mu)$ crosses the defect, and $1$ otherwise.

After each replica has been updated with standard methods~\cite{Cabibbo:1982zn,Kennedy:1985nu} (we adopt a standard $4$$:$$1$ combination of over-relaxation~\cite{Creutz:1987xi} and over-heat-bath~\cite{Brown:1987rra,Petronzio:1991gp} for this purpose), swaps among neighbouring replicas ($r$,$r+1$) are proposed via a Metropolis test with acceptance probability:
\beq
p(r,r+1) = \min\left\{1, \exp(-S_L(r\leftrightarrow r+1) + S_L(\text{no swap})) \right\}.
\eeq
Constants $c(r)$ are tuned through preliminary runs to ensure $p(r,r+1) \approx \mathrm{constant}$, so that configurations can do a random walk among different replicas and explore uniformly all boundary conditions.

Updating sweeps involving all lattice links are alternated during the Monte Carlo with \emph{hierarcic updates} involving only the links living in the neighborhood of the defect, so that the region where most of topological fluctuations are created/annihilated is updated more frequently. Finally, the periodic replica is translated by one lattice spacing in a random direction after every update in order to ensure that topological excitations are created/annihilated all over the lattice. 

\item \emph{Glueball mass computation}

We build a variational basis $\mathcal{B}_{\R^{\PC}} = \{O_i(t)\}$ of zero-momentum gauge-invariant lattice operators with the quantum numbers of the desired glueball channel $\R^{\PC}$, with $\R$ the representation of the octahedral group (which decomposes in irreducible representations of the continuum spin $J$) and $\PC$ the parity/charge conjugation. Recall that every $O_i(t)$ is computed on the periodic replica only.

Once the correlation matrix $C_{ij}(t) = \braket{O_i(t)O_j(0)}$ is obtained, we solve the Generalized Eigenvalue Problem $C_{ij}(t) v_j = C_{ij}(t^\prime) \lambda(t,t^\prime) v_j$. In particular, if we aim at determining the ground state of a certain $\R^{\PC}$ channel, it is sufficient to compute the eigenvector $\overline{v}_i$ with largest eigenvalue.

Finally, we saturate $v_i$ with $C_{ij}$ to obtain a correlator which maximizes the overlap between the vacuum and the glueball state: $C_{\mathrm{best}}(t) \equiv C_{ij}(t) \overline{v}_i \overline{v}_j \underset{t\to\infty}\sim \exp(amt)$, with $m$ the mass of the glueball state.

To properly identify the range in which $C_{\mathrm{best}}(t)$ exhibits a single-exponential decay, from which we extract the glueball mass in lattice units $am$ by means of a best fit, we look for a plateau in the \emph{effective mass}
\beq\label{eq:eff_mass}
am_{\mathrm{eff}}(t) \equiv - \log \left( \frac{ C_{\mathrm{best}}(t+1) }{ C_{\mathrm{best}}(t) } \right).
\eeq
For more details about the procedure summarized so far, we refer the reader to Refs.~\cite{BERG1983109,APE:1987ehd,TEPER1987345,Teper:1998te,Lucini:2001ej,Lucini:2004my,Blossier:2009kd,Lucini:2010nv,Bennett:2020qtj,Athenodorou:2020ani,Athenodorou:2021qvs}.
\end{itemize}

\section{Results}\label{sec:results}

\begin{figure}[!htb]
\centering
\includegraphics[scale=0.36]{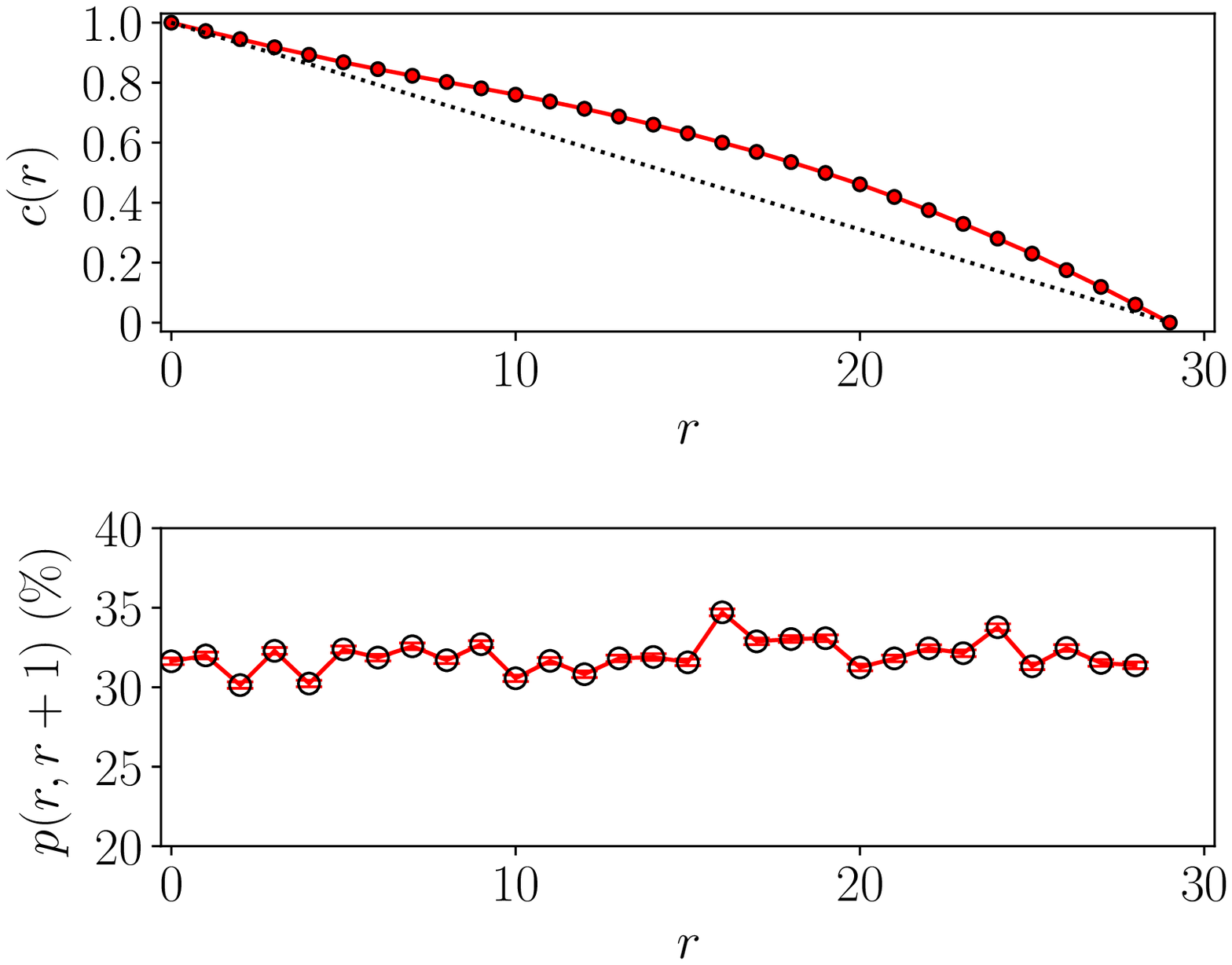}
\includegraphics[scale=0.36]{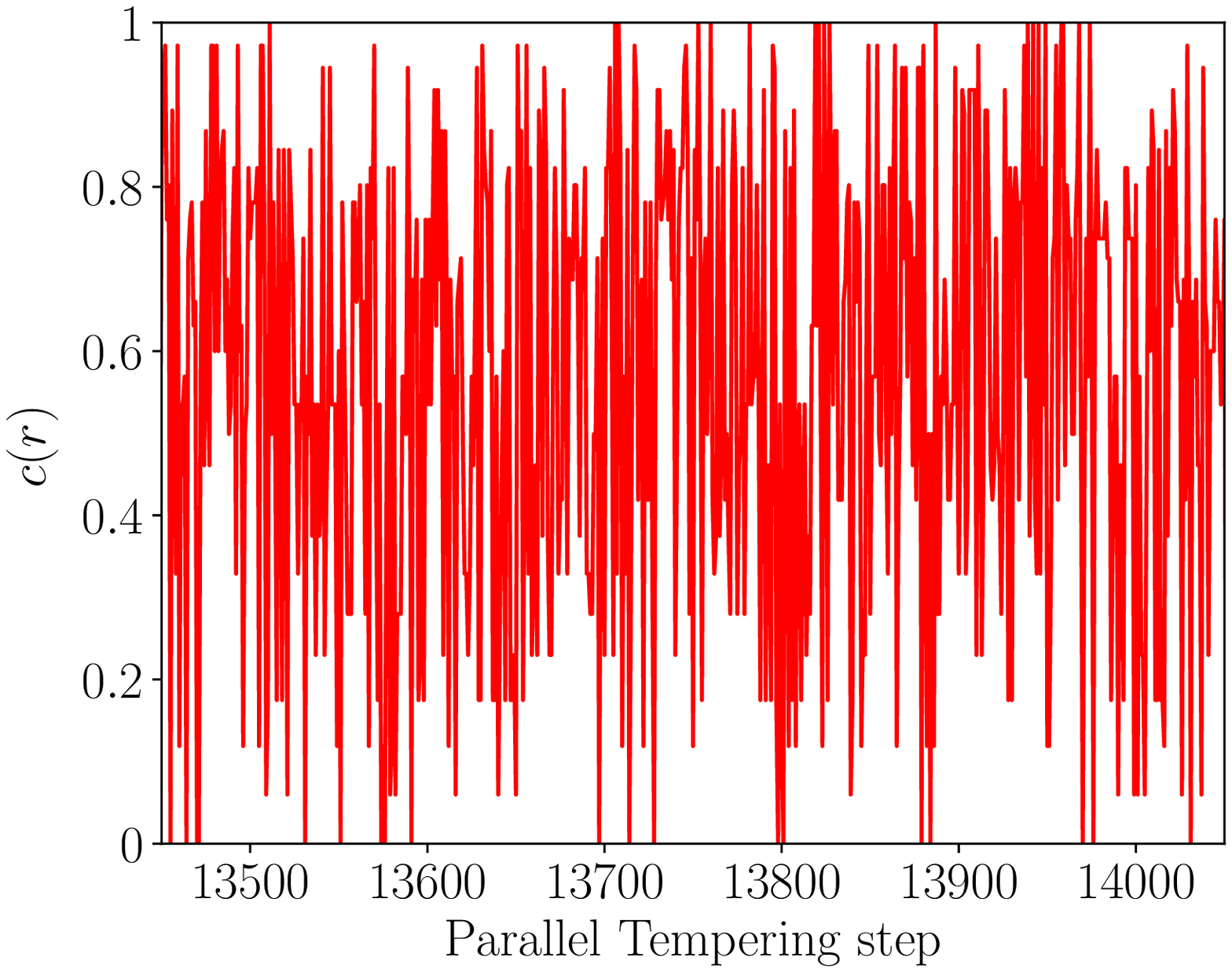}
\includegraphics[scale=0.36]{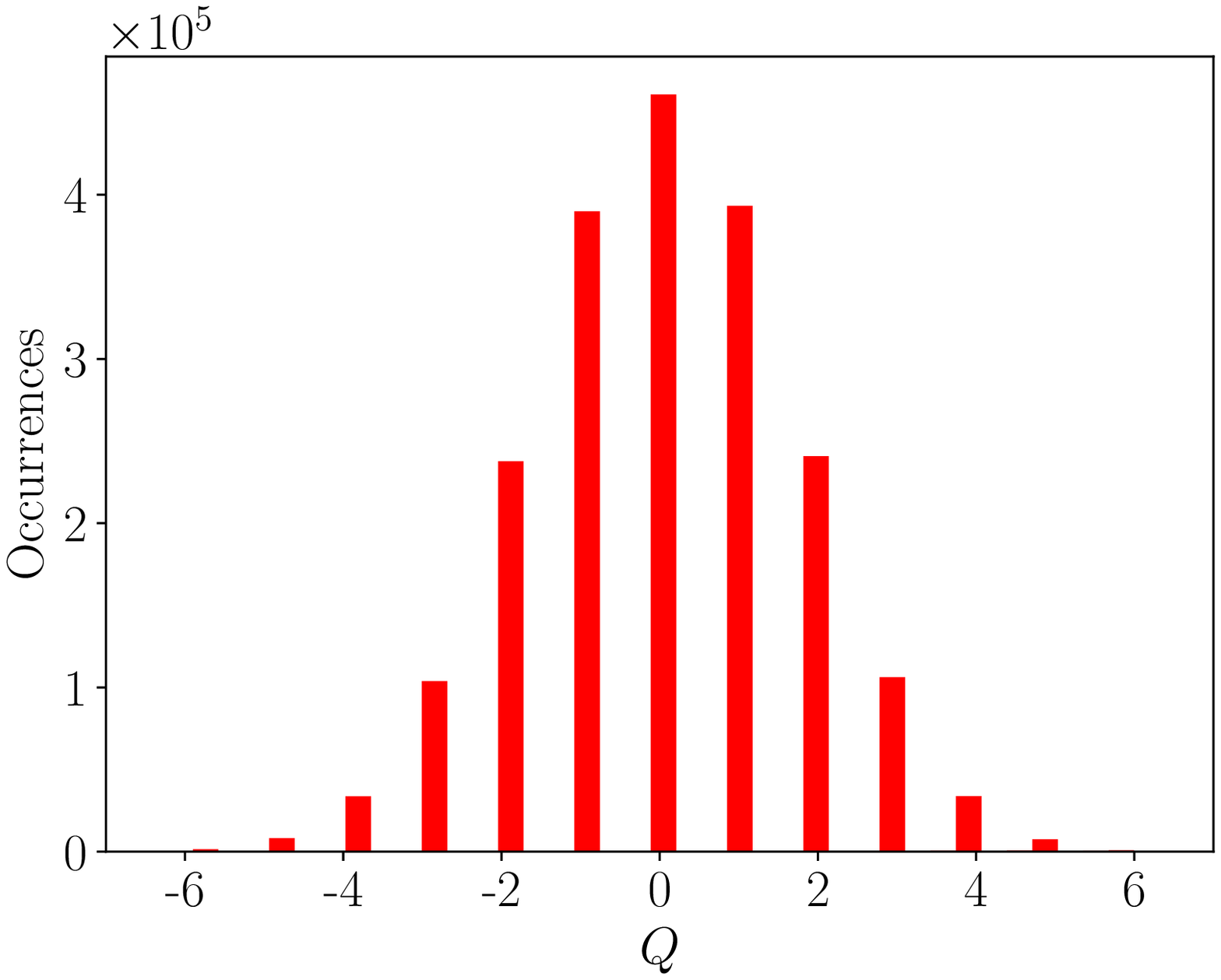}
\includegraphics[scale=0.36]{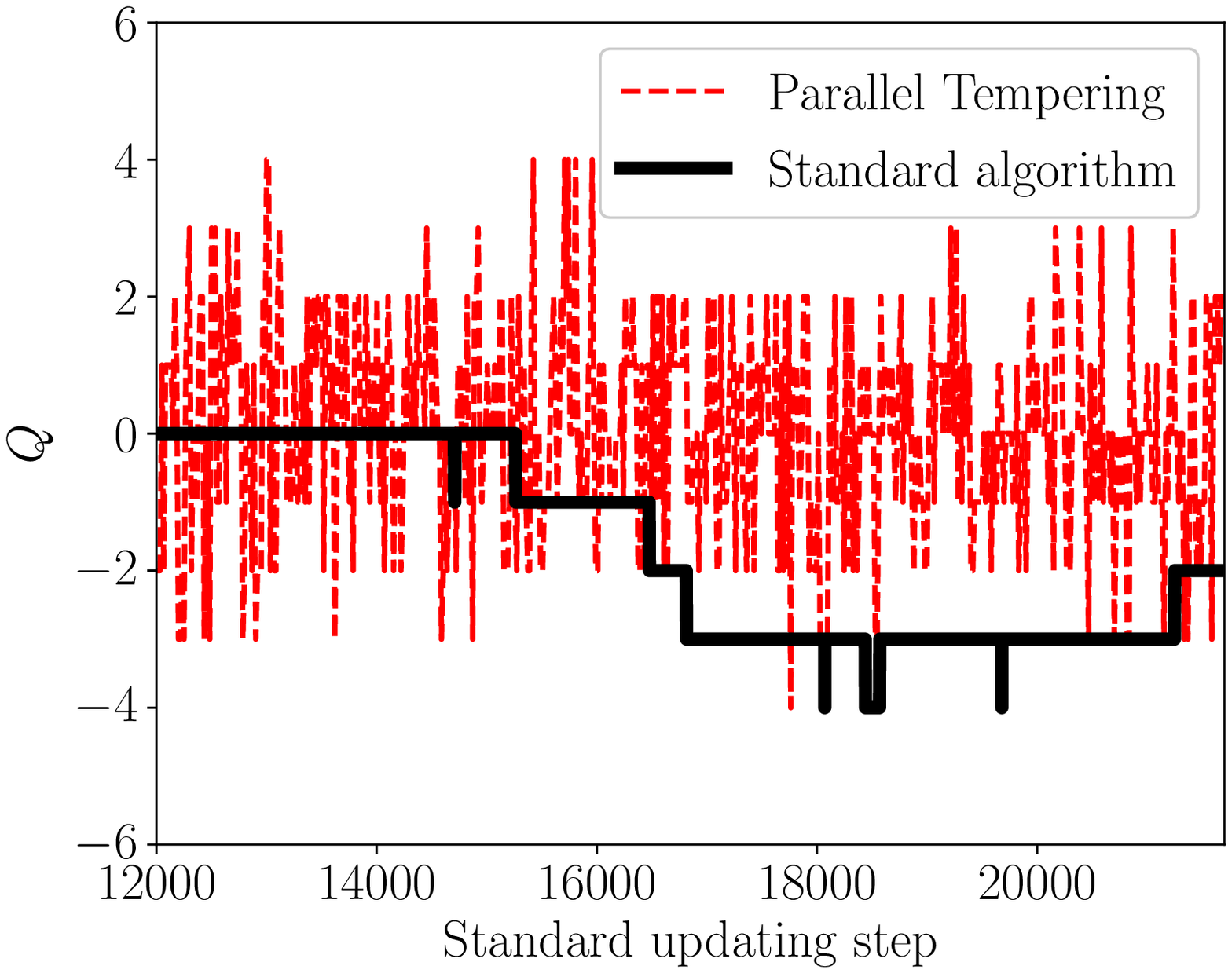}
\caption{All figures taken from Ref.~\cite{Bonanno:2022yjr}. Top left plot: behavior of $c(r)$ and $p(r,r+1)$ as a function of $r$. Top right plot: random walk of a configuration among the replicas. Bottom left plot: topological charge distribution obtained with parallel tempering. Bottom right plot: comparison of the Monte Carlo evolution of the topological charge $Q$ for the parallel tempering and the standard runs. The Monte Carlo time on the horizontal axis is reported for this plot in units of a standard updating step, i.e., the $4$$:$$1$ over-relaxation/over-heat-bath combination, for both algorithms.}
\label{fig:details_parallel_tempering}
\end{figure}

We simulated the $\SU(6)$ gauge theory on a $16^4$ lattice with $\beta=25.452$,  corresponding to a lattice spacing $a\simeq 0.0938$~fm. In Fig.~\ref{fig:details_parallel_tempering} (top left plot) we show our algorithmic setup. We employed $N_r=30$ replicas and tuned $c(r)$ to achieve a constant $\approx 30\%$ swap acceptance rate for all replicas, which ensured that each configuration uniformly explored each boundary condition (Fig.~\ref{fig:details_parallel_tempering}, top right plot). To this end, $c(r)$ has to deviate from a simple linear behavior in $r$.

The improvement obtained was noticeable: while with the standard algorithm only a handful of fluctuations of $Q$ are observed within a reasonable Monte Carlo time, parallel tempering allows to achieve many more fluctuations of $Q$ within the same simulation time, ensuring a uniform and ergodic exploration of many different topological sectors, cfr.~Fig.~\ref{fig:details_parallel_tempering} (bottom plots). We recall that we assigned an integer topological charge to each configuration through the so-called \emph{$\alpha$-rounding} of the standard \emph{clover} lattice charge computed after $30$ cooling steps. For more details on this procedure we refer the reader to, e.g., Refs.~\cite{DelDebbio:2002xa,Bonati:2016tvi,Bonanno:2020hht}.

To quantify the gain achieved by parallel tempering in terms of computational power, we computed the auto-correlation time of $Q$: while $\tau(Q)\sim 5000$ with the standard algorithm, for parallel tempering we find $\tau(Q)=92(8)$, where this number already takes into account that a single parallel tempering step requires a numerical effort which is larger by a factor of $N_r$.

We employed the generated gauge ensemble to compute glueball masses; in particular, we focused on the ground states of all $\R^{\PC}$ channels but $\mathrm{A}_1^{--}$ and $\mathrm{A}_2^{-+}$, which are found to lie above our lattice cut-off $\Lambda_{\mathrm{UV}} \sim 2/a$. We then established a correspondence between $\R$ and the representations $J$ of $\SO(3)$ as follows: $\mathrm{A}_1 \rightarrow J=0$, $\mathrm{A}_2 \rightarrow J=3$, $\mathrm{E} \rightarrow J=2$, $\mathrm{T}_1 \rightarrow J=1$, $\mathrm{T}_2 \rightarrow J=2$.

\begin{figure}[!htb]
\centering
\includegraphics[scale=0.39]{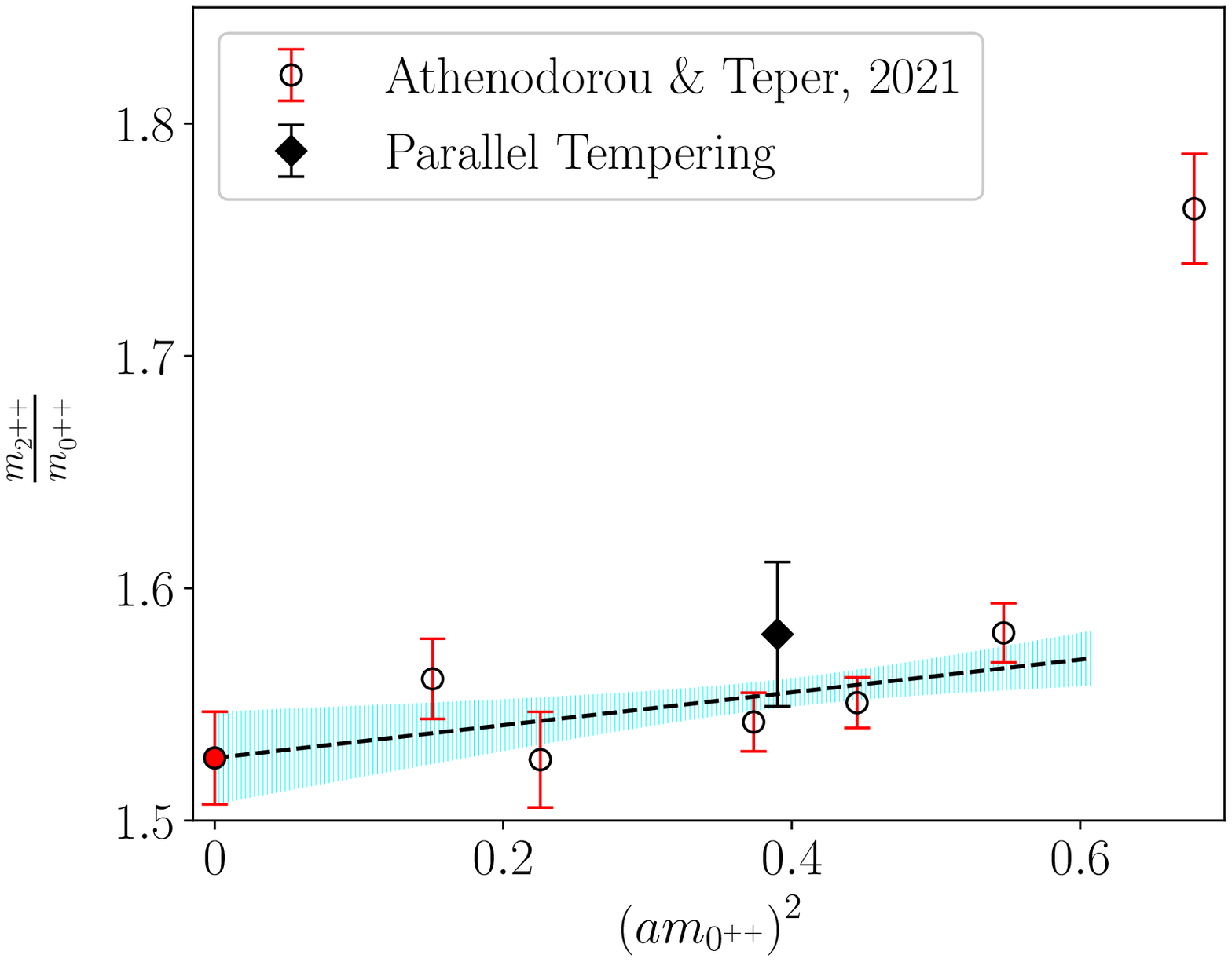}
\includegraphics[scale=0.39]{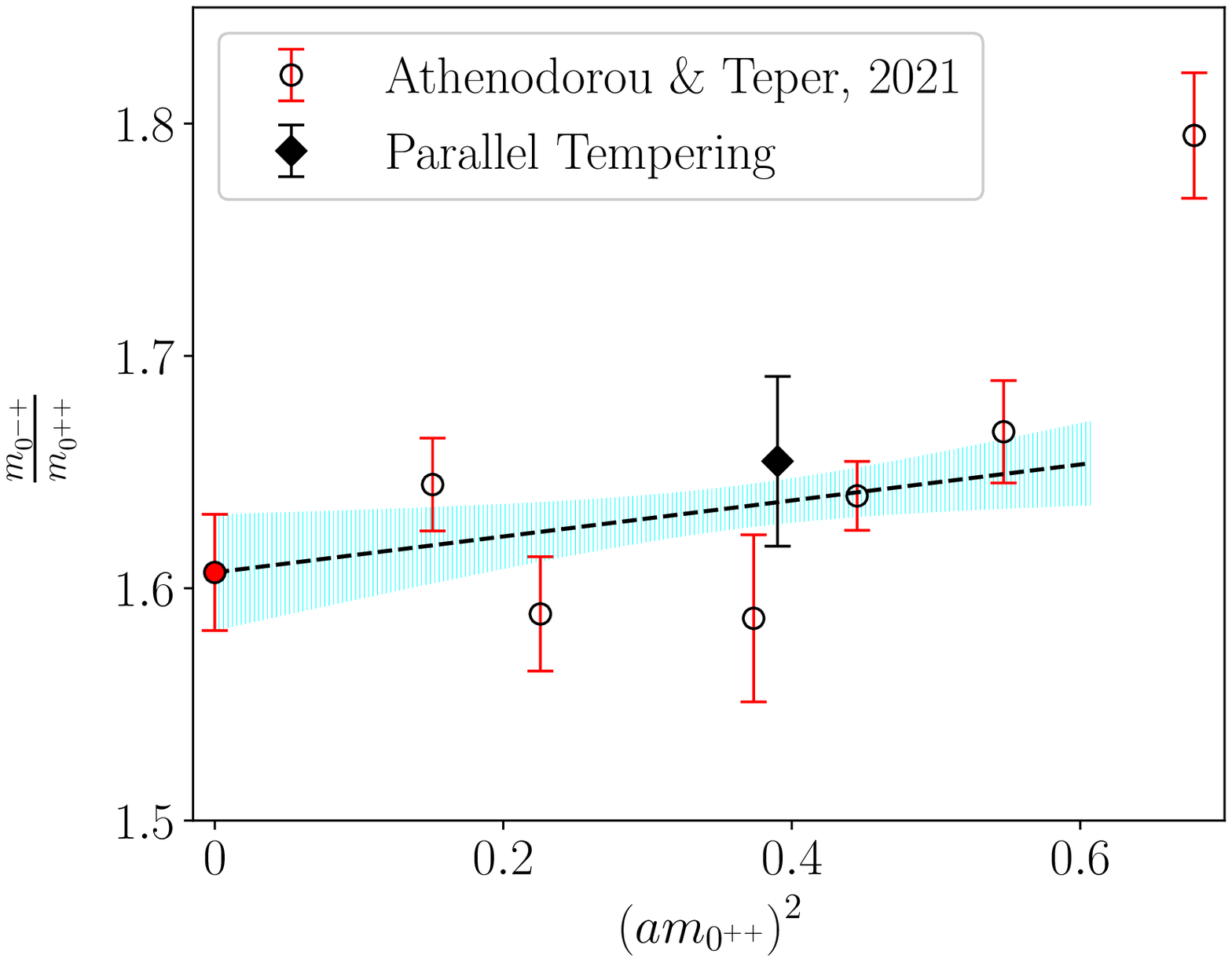}
\includegraphics[scale=0.39]{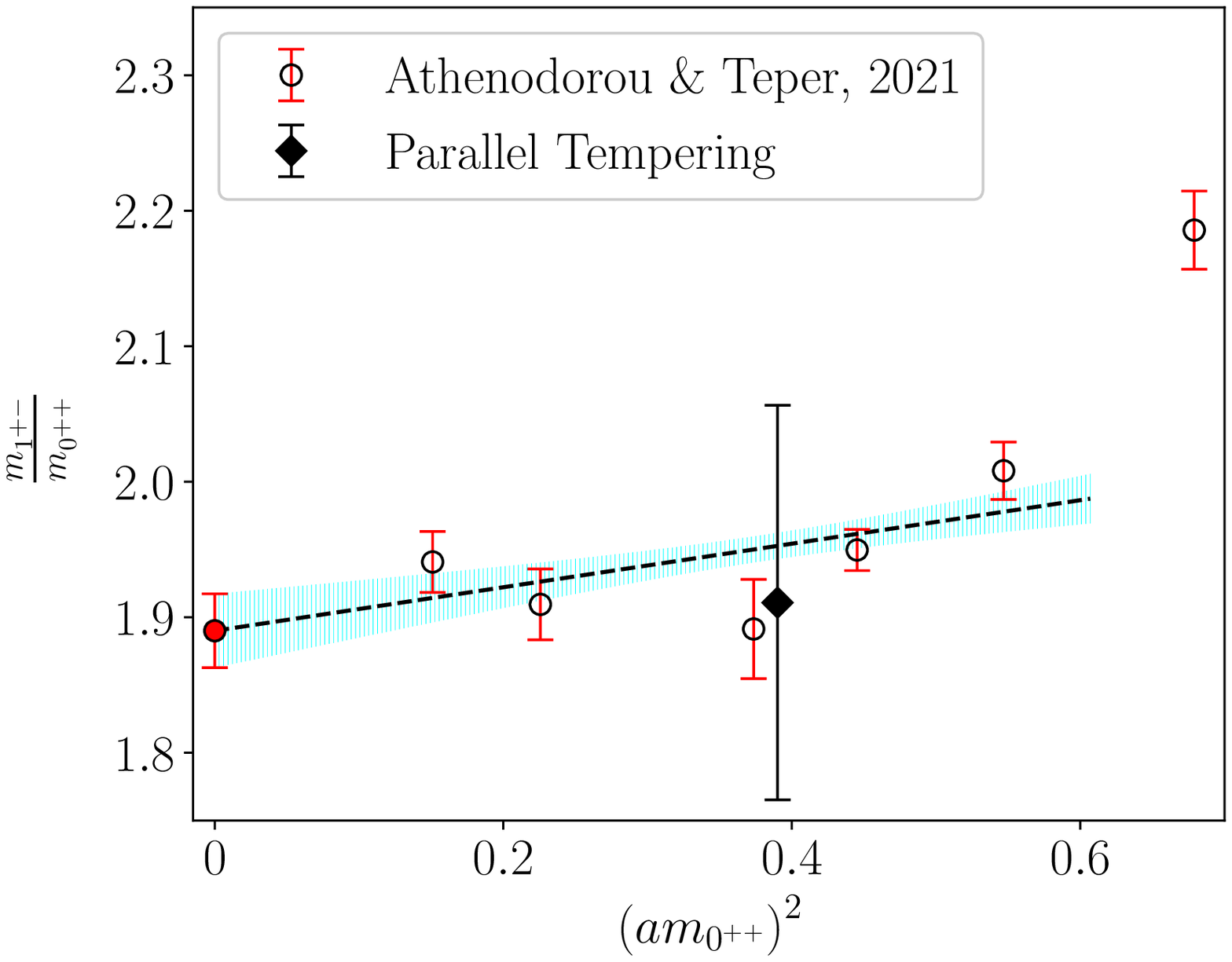}
\includegraphics[scale=0.39]{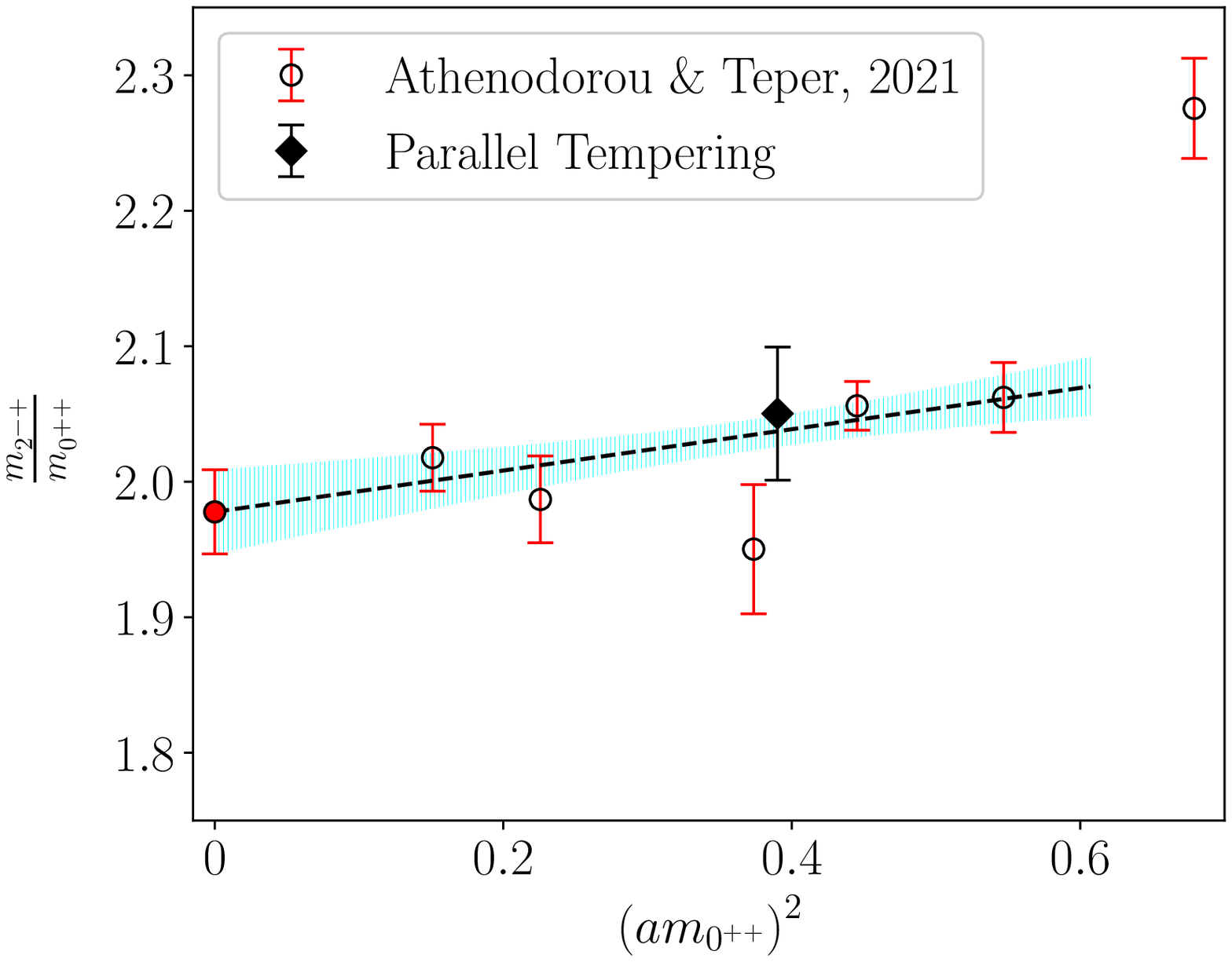}
\caption{All figures taken from Ref.~\cite{Bonanno:2022yjr}. Plots show results obtained with parallel tempering for the ratios $m_{J^{\mathrm{PC}}}/m_{0^{++}}$ for the $4$ lowest-lying states we found (diamond points). These data are compared with determinations reported in~\cite{Athenodorou:2021qvs} (empty round points). For comparison we also show the continuum extrapolation of the latter results, along with their continuum extrapolation (full round points for $a=0$) and the error on the linear best fit employed for the extrapolation (shaded areas).}
\label{fig:plot_results}
\end{figure}

To fix the scale, we employed the mass of the lightest state we found, i.e., $\mathrm{A}_1^{++}\rightarrow 0^{++}$, meaning that we considered the dimensionless ratios $m_{\mathrm{R}^{\mathrm{PC}}}/m_{\mathrm{A}_1^{++}} \rightarrow m_{J^{\mathrm{PC}}}/m_{0^{++}}$. In Figs.~\ref{fig:plot_results}, our results for the first few lightest states are compared with those obtained in Ref.~\cite{Athenodorou:2021qvs}, where masses were computed for nearly-frozen topology. It is clear that we do not detect any bias introduced by topological freezing within our percent level of precision, as our results superimpose on those of Ref.~\cite{Athenodorou:2021qvs}, even for channels with the same quantum numbers of $Q$ ($\PC=-+$).

As a final remark, we observe that our error bars are in general larger compared to those of~\cite{Athenodorou:2021qvs}. This is due to the fact that our uncertainties on glueball masses are dominated by systematic errors in the detection of the plateau of the effective mass~\eqref{eq:eff_mass}. In particular, at small times the contamination by larger-mass states is important, while for large times the effective mass is more sensitive to statistical noise. We were able to unambiguously detect a plateau for lighter states, while for heavier ones we estimated our error bars from an envelope of the quasi-plateau of $m_{\mathrm{eff}}(t)$ at large times. Nonetheless, even for heavier states we observe a substantial agreement between our findings and those of Ref.~\cite{Athenodorou:2021qvs}, thus confirming the picture clearly emerging for lighter ones.

\FloatBarrier

\section{Conclusions}\label{sec:conclusions}

In this paper we discussed the main results of~\cite{Bonanno:2022yjr}, where the first computation of glueball masses at large $N$ without any systematic effect due to topological freezing has been performed by means of the parallel tempering on boundary conditions algorithm. We explored a point of $\SU(6)$ and we determined low-lying glueball masses with $2-5\%$ precision. From the comparison of such results with those of~\cite{Athenodorou:2021qvs} we conclude that no effect of topological freezing is observed within our errors, even for those channels with the same quantum numbers of the topological charge. This comparison provides the first robust indication that determinations of glueball masses obtained from fixed-topology simulations can be trusted up to the percent level.

A possible future outlook of this work could be to tackle the problem of computing the quantity $m_2 \equiv \dfrac{d^2 m(\theta)}{d\theta^2}\bigg\vert_{\theta=0}$, where $\theta$ is the dimensionless parameter coupling $Q$ to the Yang--Mills action, being any correction due to fixed topology to a glueball mass $m$ proportional to this quantity~\cite{Brower:2003yx}. A direct computation of $m_2$ is hard, especially at large $N$ and for heavier states, as shown in~\cite{DelDebbio:2004xh}, because this quantity is affected by large statistical noise. Possible improvements are expected to be obtained combining imaginary-$\theta$ simulations~\cite{Bonati:2015sqt,Bonati:2016tvi,Bonanno:2018xtd,Berni:2019bch,Bonanno:2020hht} with the parallel tempering algorithm here discussed, as already shown in~\cite{Bonanno:2020hht}, where the combination of these two methods has been employed to improve the computation of higher-order terms in the $\theta$-expansion of the vacuum energy around $\theta=0$ at large $N$.

\section*{Acknowledgements}
The authors thank A.~Athenodorou and T.~DeGrand for useful discussions.

C.~B.~acknowledges the support of the Italian Ministry of Education, University and Research under the project PRIN 2017E44HRF, ``Low dimensional quantum systems: theory, experiments and simulations''.

The work of B.~L.~has been supported in part by the STFC Consolidated Grants No. ST/P00055X/1 and No. ST/T000813/1. B.~L.~received funding from the European Research Council (ERC) under the European Union’s Horizon 2020 research and innovation program under Grant Agreement No.~813942. The work of B.~L.~is further supported in part by the Royal Society WolfsonResearch Merit Award No.~WM170010 and by the Leverhulme Trust Research Fellowship No.~RF-2020-4619.

The work of D.~V.~is partly supported by the Simons Foundation under the program ``Targeted Grants to Institutes'' awarded to the Hamilton Mathematics Institute.

Numerical simulations have been performed on the \texttt{MARCONI} machine at CINECA, based on the agreement between INFN and CINECA, under project INF21\_npqcd. Numerical analyses have been performed on the Swansea University \texttt{SUNBIRD} (part of the Supercomputing Wales project) and \texttt{AccelerateAI} A100 GPU system, which are part funded by the European Regional Development Fund (ERDF) via Welsh Government.

\providecommand{\href}[2]{#2}\begingroup\raggedright\endgroup


\begin{thebibliography}{10}
	
	\bibitem{TOTEM:2020zzr}
	{\scshape TOTEM, D0} collaboration, V.~M. Abazov et~al., \emph{{Odderon
			Exchange from Elastic Scattering Differences between $pp$ and $p \bar{p}$
			Data at 1.96~TeV and from pp Forward Scattering Measurements}},
	\href{https://doi.org/10.1103/PhysRevLett.127.062003}{\emph{Phys. Rev. Lett.}
		{\bfseries 127} (2021) 062003}
	[\href{https://arxiv.org/abs/2012.03981}{{\ttfamily 2012.03981}}].
	
	\bibitem{Lucini:2001ej}
	B.~Lucini and M.~Teper, \emph{{SU($N$) gauge theories in four-dimensions:
			Exploring the approach to $N = \infty$}},
	\href{https://doi.org/10.1088/1126-6708/2001/06/050}{\emph{JHEP} {\bfseries
			06} (2001) 050} [\href{https://arxiv.org/abs/hep-lat/0103027}{{\ttfamily
			hep-lat/0103027}}].
	
	\bibitem{Lucini:2004my}
	B.~Lucini, M.~Teper and U.~Wenger, \emph{{Glueballs and $k$-strings in SU($N$)
			gauge theories: Calculations with improved operators}},
	\href{https://doi.org/10.1088/1126-6708/2004/06/012}{\emph{JHEP} {\bfseries
			06} (2004) 012} [\href{https://arxiv.org/abs/hep-lat/0404008}{{\ttfamily
			hep-lat/0404008}}].
	
	\bibitem{Lucini:2010nv}
	B.~Lucini, A.~Rago and E.~Rinaldi, \emph{{Glueball masses in the large $N$
			limit}}, \href{https://doi.org/10.1007/JHEP08(2010)119}{\emph{JHEP}
		{\bfseries 08} (2010) 119} [\href{https://arxiv.org/abs/1007.3879}{{\ttfamily
			1007.3879}}].
	
	\bibitem{Hong:2017suj}
	D.~K. Hong, J.-W. Lee, B.~Lucini, M.~Piai and D.~Vadacchino, \emph{{Casimir
			scaling and Yang\textendash{}Mills glueballs}},
	\href{https://doi.org/10.1016/j.physletb.2017.10.050}{\emph{Phys. Lett. B}
		{\bfseries 775} (2017) 89}
	[\href{https://arxiv.org/abs/1705.00286}{{\ttfamily 1705.00286}}].
	
	\bibitem{Bennett:2020hqd}
	E.~Bennett, J.~Holligan, D.~K. Hong, J.-W. Lee, C.~J.~D. Lin, B.~Lucini et~al.,
	\emph{{Color dependence of tensor and scalar glueball masses in Yang-Mills
			theories}}, \href{https://doi.org/10.1103/PhysRevD.102.011501}{\emph{Phys.
			Rev. D} {\bfseries 102} (2020) 011501}
	[\href{https://arxiv.org/abs/2004.11063}{{\ttfamily 2004.11063}}].
	
	\bibitem{Bennett:2020qtj}
	E.~Bennett, J.~Holligan, D.~K. Hong, J.-W. Lee, C.~J.~D. Lin, B.~Lucini et~al.,
	\emph{{Glueballs and strings in $Sp(2N)$ Yang-Mills theories}},
	\href{https://doi.org/10.1103/PhysRevD.103.054509}{\emph{Phys. Rev. D}
		{\bfseries 103} (2021) 054509}
	[\href{https://arxiv.org/abs/2010.15781}{{\ttfamily 2010.15781}}].
	
	\bibitem{Athenodorou:2020ani}
	A.~Athenodorou and M.~Teper, \emph{{The glueball spectrum of SU($3$) gauge
			theory in $3 + 1$ dimensions}},
	\href{https://doi.org/10.1007/JHEP11(2020)172}{\emph{JHEP} {\bfseries 11}
		(2020) 172} [\href{https://arxiv.org/abs/2007.06422}{{\ttfamily
			2007.06422}}].
	
	\bibitem{Athenodorou:2021qvs}
	A.~Athenodorou and M.~Teper, \emph{{SU($N$) gauge theories in $3 + 1$
			dimensions: glueball spectrum, string tensions and topology}},
	\href{https://doi.org/10.1007/JHEP12(2021)082}{\emph{JHEP} {\bfseries 12}
		(2021) 082} [\href{https://arxiv.org/abs/2106.00364}{{\ttfamily
			2106.00364}}].
	
	\bibitem{BERG1983109}
	B.~Berg and A.~Billoire, \emph{Glueball spectroscopy in $4d$ su($3$) lattice
		gauge theory (i)},
	\href{https://doi.org/https://doi.org/10.1016/0550-3213(83)90620-X}{\emph{Nuclear
			Physics B} {\bfseries 221} (1983) 109}.
	
	\bibitem{APE:1987ehd}
	{\scshape APE} collaboration, M.~Albanese et~al., \emph{{Glueball Masses and
			String Tension in Lattice QCD}},
	\href{https://doi.org/10.1016/0370-2693(87)91160-9}{\emph{Phys. Lett. B}
		{\bfseries 192} (1987) 163}.
	
	\bibitem{TEPER1987345}
	M.~Teper, \emph{An improved method for lattice glueball calculations},
	\href{https://doi.org/https://doi.org/10.1016/0370-2693(87)90976-2}{\emph{Physics
			Letters B} {\bfseries 183} (1987) 345}.
	
	\bibitem{Morningstar:1997ff}
	C.~J. Morningstar and M.~J. Peardon, \emph{{Efficient glueball simulations on
			anisotropic lattices}},
	\href{https://doi.org/10.1103/PhysRevD.56.4043}{\emph{Phys. Rev. D}
		{\bfseries 56} (1997) 4043}
	[\href{https://arxiv.org/abs/hep-lat/9704011}{{\ttfamily hep-lat/9704011}}].
	
	\bibitem{Teper:1998te}
	M.~J. Teper, \emph{{SU($N$) gauge theories in $(2+1)$-dimensions}},
	\href{https://doi.org/10.1103/PhysRevD.59.014512}{\emph{Phys. Rev. D}
		{\bfseries 59} (1999) 014512}
	[\href{https://arxiv.org/abs/hep-lat/9804008}{{\ttfamily hep-lat/9804008}}].
	
	\bibitem{Morningstar:1999rf}
	C.~J. Morningstar and M.~J. Peardon, \emph{{The Glueball spectrum from an
			anisotropic lattice study}},
	\href{https://doi.org/10.1103/PhysRevD.60.034509}{\emph{Phys. Rev. D}
		{\bfseries 60} (1999) 034509}
	[\href{https://arxiv.org/abs/hep-lat/9901004}{{\ttfamily hep-lat/9901004}}].
	
	\bibitem{Blossier:2009kd}
	B.~Blossier, M.~Della~Morte, G.~von Hippel, T.~Mendes and R.~Sommer, \emph{{On
			the generalized eigenvalue method for energies and matrix elements in lattice
			field theory}},
	\href{https://doi.org/10.1088/1126-6708/2009/04/094}{\emph{JHEP} {\bfseries
			04} (2009) 094} [\href{https://arxiv.org/abs/0902.1265}{{\ttfamily
			0902.1265}}].
	
	\bibitem{Alles:1996vn}
	B.~Alles, G.~Boyd, M.~D'Elia, A.~Di~Giacomo and E.~Vicari, \emph{{Hybrid Monte
			Carlo and topological modes of full QCD}},
	\href{https://doi.org/10.1016/S0370-2693(96)01247-6}{\emph{Phys. Lett.}
		{\bfseries B389} (1996) 107}
	[\href{https://arxiv.org/abs/hep-lat/9607049}{{\ttfamily hep-lat/9607049}}].
	
	\bibitem{DelDebbio:2004xh}
	L.~Del~Debbio, G.~M. Manca and E.~Vicari, \emph{{Critical slowing down of
			topological modes}},
	\href{https://doi.org/10.1016/j.physletb.2004.05.038}{\emph{Phys. Lett. B}
		{\bfseries 594} (2004) 315}
	[\href{https://arxiv.org/abs/hep-lat/0403001}{{\ttfamily hep-lat/0403001}}].
	
	\bibitem{Schaefer:2010hu}
	{\scshape ALPHA} collaboration, S.~Schaefer, R.~Sommer and F.~Virotta,
	\emph{{Critical slowing down and error analysis in lattice QCD simulations}},
	\href{https://doi.org/10.1016/j.nuclphysb.2010.11.020}{\emph{Nucl. Phys.}
		{\bfseries B845} (2011) 93}
	[\href{https://arxiv.org/abs/1009.5228}{{\ttfamily 1009.5228}}].
	
	\bibitem{Lucini:2004yh}
	B.~Lucini, M.~Teper and U.~Wenger, \emph{{Topology of $\mathrm{SU}(N)$ gauge
			theories at $T \simeq 0$ and $T \simeq T_c$}},
	\href{https://doi.org/10.1016/j.nuclphysb.2005.02.037}{\emph{Nucl. Phys.}
		{\bfseries B715} (2005) 461}
	[\href{https://arxiv.org/abs/hep-lat/0401028}{{\ttfamily hep-lat/0401028}}].
	
	\bibitem{DelDebbio:2006yuf}
	L.~Del~Debbio, G.~M. Manca, H.~Panagopoulos, A.~Skouroupathis and E.~Vicari,
	\emph{{Theta-dependence of the spectrum of $SU(N)$ gauge theories}},
	\href{https://doi.org/10.1088/1126-6708/2006/06/005}{\emph{JHEP} {\bfseries
			06} (2006) 005} [\href{https://arxiv.org/abs/hep-th/0603041}{{\ttfamily
			hep-th/0603041}}].
	
	\bibitem{Bonati:2016tvi}
	C.~Bonati, M.~D'Elia, P.~Rossi and E.~Vicari, \emph{{$\theta$ dependence of 4D
			$SU(N)$ gauge theories in the large-$N$ limit}},
	\href{https://doi.org/10.1103/PhysRevD.94.085017}{\emph{Phys. Rev.}
		{\bfseries D94} (2016) 085017}
	[\href{https://arxiv.org/abs/1607.06360}{{\ttfamily 1607.06360}}].
	
	\bibitem{Bonati:2017woi}
	C.~Bonati and M.~D'Elia, \emph{{Topological critical slowing down: variations
			on a toy model}},
	\href{https://doi.org/10.1103/PhysRevE.98.013308}{\emph{Phys. Rev.}
		{\bfseries E98} (2018) 013308}
	[\href{https://arxiv.org/abs/1709.10034}{{\ttfamily 1709.10034}}].
	
	\bibitem{Bonanno:2018xtd}
	C.~Bonanno, C.~Bonati and M.~D'Elia, \emph{{Topological properties of
			$CP^{N-1}$ models in the large-$N$ limit}},
	\href{https://doi.org/10.1007/JHEP01(2019)003}{\emph{JHEP} {\bfseries 01}
		(2019) 003} [\href{https://arxiv.org/abs/1807.11357}{{\ttfamily
			1807.11357}}].
	
	\bibitem{Berni:2019bch}
	M.~Berni, C.~Bonanno and M.~D'Elia, \emph{{Large-$N$ expansion and
			$\theta$-dependence of $2d$ $CP^{N-1}$ models beyond the leading order}},
	\href{https://doi.org/10.1103/PhysRevD.100.114509}{\emph{Phys. Rev.}
		{\bfseries D100} (2019) 114509}
	[\href{https://arxiv.org/abs/1911.03384}{{\ttfamily 1911.03384}}].
	
	\bibitem{Berni:2020ebn}
	M.~Berni, C.~Bonanno and M.~D'Elia, \emph{{$\theta$-dependence in the small-$N$
			limit of $2d$ $CP^{N-1}$ models}},
	\href{https://doi.org/10.1103/PhysRevD.102.114519}{\emph{Phys. Rev. D}
		{\bfseries 102} (2020) 114519}
	[\href{https://arxiv.org/abs/2009.14056}{{\ttfamily 2009.14056}}].
	
	\bibitem{Bennett:2022ftz}
	Bennett, D.~K. Hong, J.-W. Lee, C.~J.~D. Lin, B.~Lucini, M.~Piai et~al.,
	\emph{{$Sp(2N)$ Yang-Mills theories on the lattice: scale setting and
			topology}},  \href{https://arxiv.org/abs/2205.09364}{{\ttfamily 2205.09364}}.
	
	\bibitem{Bennett:2022gdz}
	Bennett, D.~K. Hong, J.-W. Lee, C.~J.~D. Lin, B.~Lucini, M.~Piai et~al.,
	\emph{{Color dependence of the topological susceptibility in Yang-Mills
			theories}},  \href{https://arxiv.org/abs/2205.09254}{{\ttfamily 2205.09254}}.
	
	\bibitem{Brower:2003yx}
	R.~Brower, S.~Chandrasekharan, J.~W. Negele and U.~J. Wiese, \emph{{QCD at
			fixed topology}},
	\href{https://doi.org/10.1016/S0370-2693(03)00369-1}{\emph{Phys. Lett. B}
		{\bfseries 560} (2003) 64}
	[\href{https://arxiv.org/abs/hep-lat/0302005}{{\ttfamily hep-lat/0302005}}].
	
	\bibitem{Bonanno:2022yjr}
	C.~Bonanno, M.~D'Elia, B.~Lucini and D.~Vadacchino, \emph{{Towards glueball
			masses of large-$N$ $\mathrm{SU}(N)$ pure-gauge theories without topological
			freezing}}, \href{https://doi.org/10.1016/j.physletb.2022.137281}{\emph{Phys.
			Lett. B} {\bfseries 833} (2022) 137281}
	[\href{https://arxiv.org/abs/2205.06190}{{\ttfamily 2205.06190}}].
	
	\bibitem{Hasenbusch:2017unr}
	M.~Hasenbusch, \emph{{Fighting topological freezing in the two-dimensional
			$CP^{N-1}$ model}},
	\href{https://doi.org/10.1103/PhysRevD.96.054504}{\emph{Phys. Rev. D}
		{\bfseries 96} (2017) 054504}
	[\href{https://arxiv.org/abs/1706.04443}{{\ttfamily 1706.04443}}].
	
	\bibitem{Bonanno:2020hht}
	C.~Bonanno, C.~Bonati and M.~D'Elia, \emph{{Large-$N$ $SU(N)$ Yang-Mills
			theories with milder topological freezing}},
	\href{https://doi.org/10.1007/JHEP03(2021)111}{\emph{JHEP} {\bfseries 03}
		(2021) 111} [\href{https://arxiv.org/abs/2012.14000}{{\ttfamily
			2012.14000}}].
	
	\bibitem{Cabibbo:1982zn}
	N.~Cabibbo and E.~Marinari, \emph{{A New Method for Updating SU(N) Matrices in
			Computer Simulations of Gauge Theories}},
	\href{https://doi.org/10.1016/0370-2693(82)90696-7}{\emph{Phys. Lett. B}
		{\bfseries 119} (1982) 387}.
	
	\bibitem{Kennedy:1985nu}
	A.~D. Kennedy and B.~J. Pendleton, \emph{{Improved Heat Bath Method for Monte
			Carlo Calculations in Lattice Gauge Theories}},
	\href{https://doi.org/10.1016/0370-2693(85)91632-6}{\emph{Phys. Lett.}
		{\bfseries 156B} (1985) 393}.
	
	\bibitem{Creutz:1987xi}
	M.~Creutz, \emph{{Overrelaxation and Monte Carlo Simulation}},
	\href{https://doi.org/10.1103/PhysRevD.36.515}{\emph{Phys. Rev.} {\bfseries
			D36} (1987) 515}.
	
	\bibitem{Brown:1987rra}
	F.~R. Brown and T.~J. Woch, \emph{{Overrelaxed Heat Bath and Metropolis
			Algorithms for Accelerating Pure Gauge Monte Carlo Calculations}},
	\href{https://doi.org/10.1103/PhysRevLett.58.2394}{\emph{Phys. Rev. Lett.}
		{\bfseries 58} (1987) 2394}.
	
	\bibitem{Petronzio:1991gp}
	R.~Petronzio and E.~Vicari, \emph{{An Overheat bath algorithm for lattice gauge
			theories}}, \href{https://doi.org/10.1016/0370-2693(91)91183-V}{\emph{Phys.
			Lett. B} {\bfseries 254} (1991) 444}.
	
	\bibitem{DelDebbio:2002xa}
	L.~Del~Debbio, H.~Panagopoulos and E.~Vicari, \emph{{theta dependence of SU(N)
			gauge theories}},
	\href{https://doi.org/10.1088/1126-6708/2002/08/044}{\emph{JHEP} {\bfseries
			08} (2002) 044} [\href{https://arxiv.org/abs/hep-th/0204125}{{\ttfamily
			hep-th/0204125}}].
	
	\bibitem{Bonati:2015sqt}
	C.~Bonati, M.~D'Elia and A.~Scapellato, \emph{{$\theta$ dependence in $SU(3)$
			Yang-Mills theory from analytic continuation}},
	\href{https://doi.org/10.1103/PhysRevD.93.025028}{\emph{Phys. Rev.}
		{\bfseries D93} (2016) 025028}
	[\href{https://arxiv.org/abs/1512.01544}{{\ttfamily 1512.01544}}].
	
\end{thebibliography}
\end{document}